\documentclass[a4paper,11pt]{article}
\usepackage{pos}
\usepackage{amsmath,graphicx,multirow,tabularx,physics}
\usepackage{davitex,slashed,listings}

\definecolor{airforceblue}{rgb}{0.36,0.54,0.66}
\definecolor{burgundy}{rgb}{0.5,0.0,0.13}
\definecolor{blue-violet}{rgb}{0.54,0.17,0.89}
\newcommand{\simulat}{\texttt{SIMULATeQCD}}

\title{HotQCD on Multi-GPU Systems}
\ShortTitle{HotQCD on Multi-GPU Systems}

\author[a]{L. Altenkort}
\author*[a]{D. Bollweg}
\author[a]{D. A. Clarke}
\author[a]{O. Kaczmarek}
\author[a,b]{L. Mazur}
\author[a]{C. Schmidt}
\author[a,c]{P. Scior}
\author[a,d]{Hai-Tao Shu}

\affiliation[a]{Fakult\"at f\"ur Physik, Universit\"at Bielefeld,\\
Bielefeld, Germany} 

\affiliation[b]{Paderborn Center for Parallel Computing, Paderborn University,\\
Paderborn, Germany}

\affiliation[c]{Physics Department, Brookhaven National Laboratory,\\
Upton, New York, United States}

\affiliation[d]{Institut f\"ur Theoretische Physik, Universit\"at  Regensburg,\\
Regensburg, Germany}

\emailAdd{altenkort@physik.uni-bielefeld.de}
\emailAdd{dennis.bollweg@uni-bielefeld.de}
\emailAdd{dclarke@physik.uni-bielefeld.de}
\emailAdd{okacz@physik.uni-bielefeld.de}
\emailAdd{lukas.mazur@uni-paderborn.de}
\emailAdd{schmidt@physik.uni-bielefeld.de}
\emailAdd{pscior@bnl.gov}
\emailAdd{hai-tao.shu@ur.de}

\abstract{We present $\simulat$, HotQCD's software for performing lattice QCD calculations 
on GPUs. Started in late 2017 and intended as a full replacement of the previous single 
GPU lattice QCD code used by the HotQCD collaboration, our software has been 
developed into an extensive framework for lattice QCD calculations distributed on multiple 
GPUs over many compute nodes. The code is built on C++, CUDA, and MPI and leverages 
modern C++ language features to provide high-level data structures, objects, and 
algorithms that allow users to express lattice QCD calculations in an intuitive way 
without sacrificing performance. Implemented algorithms range from gradient flow, 
correlator measurements, and mixed precision conjugate gradient solvers all the way to 
full HISQ gauge field configuration generation using RHMC. After successful deployment 
in large-scale computing projects, we want to share the result of our efforts with the 
lattice QCD community by making it publicly available. 
In these proceedings, we will present some of the key features of our 
code, demonstrate its ease of use, and show benchmarks of performance critical kernels 
on state-of-the-art supercomputers.}

\FullConference{%
 The 38th International Symposium on Lattice Field Theory, LATTICE2021
  26th-30th July, 2021
  Zoom/Gather@Massachusetts Institute of Technology
}

\begin{document}
\maketitle

\section{Introduction}\label{sec:intro}

Understanding the structure of the QCD phase diagram
is a rich area of high energy physics research with important phenomenological
implications for heavy-ion collisions, early universe physics, and compact stars. 
Lattice calculations allow us to access regions of the diagram
at large temperatures and zero-to-moderate chemical potential. 
Of special interest is research into the nature of the QCD critical
point, and there is much effort nowadays trying to tackle this problem from the
perspective of the chiral limit~\cite{karsch_critical_2019}.  

Such investigations demand high-performance, highly parallelized code, as
increasing the lattice size, computing higher-order cumulants, and decreasing the
light quark mass, especially with an eye toward continuum-limit extrapolations,
progressively push the limits of available computational power. 
Current investigations toward the chiral limit aim at and below light quark masses
of $m_l=m_s/160$~\cite{ding_chiral_2019,Clarke:2020htu}, with computational cost increasing
dramatically as $m_l$ is lowered. This pushes us to develop high-performance 
GPU code that can efficiently
generate configurations with dynamical quarks using the HISQ action~\cite{follana_highly_2007}.
Studies of two-point functions defined in the Euclidean time direction, such
as hadron correlators or the gluonic color-electric correlator, demand lattices with a large time extension and
correspondingly even larger spatial extensions, with modern lattices 
reaching sizes as large as $144^3\times36$~\cite{altenkort_heavy_2021}. These
large gauge fields cannot always be accommodated by a single GPU, and hence
it is important to be able to split the lattice among multiple GPUs.

The HotQCD collaboration's lattice studies take place within the context of 
large-scale computing
projects on several Top500 systems, including Summit (OLCF), Marconi100
(CINECA), JUWELS (JSC), and Piz Daint (CSCS), in addition to running on
Bielefeld's local GPU cluster. Hence it is also important that we have flexible
code that can function on different architectures. Looking ahead a bit, it is
also important that this code works for different GPU manufacturers; for example
Frontier, the successor to Summit, will use AMD cards instead of NVIDIA.
This need for flexibility spurs us to write highly modular code so that
one can easily expand, adapt, and maintain parts of the code close to the 
hardware without needing to make any changes to higher-level classes and 
structures. Along this vein, we would like to
leverage features of modern C++ to allow physicists with
intermediate C++ knowledge to easily and intuitively carry out lattice
calculations.

These are the goals we had in mind when developing our (Si)mple, (Mu)lti-GPU
(Lat)ice code for (QCD) calculations, which we stylize as $\simulat$. 
$\simulat$ is the successor to HotQCD's previous single-GPU code, the
\texttt{BielefeldGPUCode}. 

The increasing availability of multi-GPU systems motivated the Bielefeld lattice group in 2017
to add multi-GPU functionality; this presented an opportunity to develop a modern, future-proof, 
multi-GPU framework for lattice QCD calculations with completely revised implementations of the basic routines.
In these proceedings we wish to share our progress writing this code,
discussing its design and structure (Section~\ref{sec:design}), and 
showcasing its performance (Section~\ref{sec:perform}).

\section{Design strategy and available modules}\label{sec:design}

$\simulat$ is a multi-GPU, multi-node lattice code written in C++ and utilizing the
OOP paradigm and modern C++ features. It was originally written to run on
multiple NVIDIA GPUs using CUDA, but it also supports AMD GPUs through HIP. It is also
possible to run the code on multiple CPUs. In the following we discuss key ideas
guiding the code's design and mention some of the tools already available for
state-of-the-art lattice calculations. A more detailed discussion of
$\simulat$\footnote{In this reference the code is referred to as the
\texttt{ParallelGPUCode}, which was its working title at that time.}
can be found in Ref.~\cite{mazur_topological_2021}. 

The development of $\simulat$ targets calculations of QCD at high temperature and density, particularly in the
regimes and limits listed in Section~\ref{sec:intro}. With these applications in
mind, we designed the code such that it
\begin{enumerate}
  \item is high-performant;
  \item works efficiently on multiple GPUs and nodes;
  \item is flexible to changing architecture and hardware;
  \item is easy to use for lattice practitioners with intermediate C++
        knowledge; and
  \item contains tools needed for calculating observables of interest to hot and dense QCD.
\end{enumerate}

\begin{figure}
\centering
\includegraphics[width=0.35\textwidth]{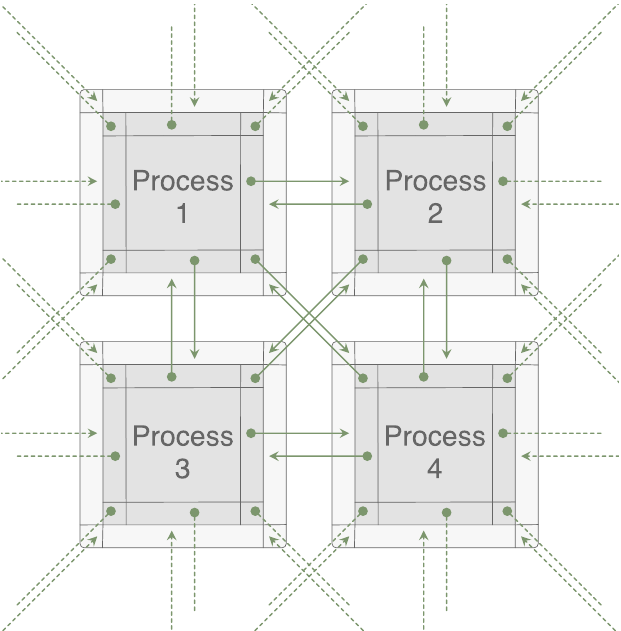}
\hspace{5mm}
\includegraphics[width=0.55\textwidth]{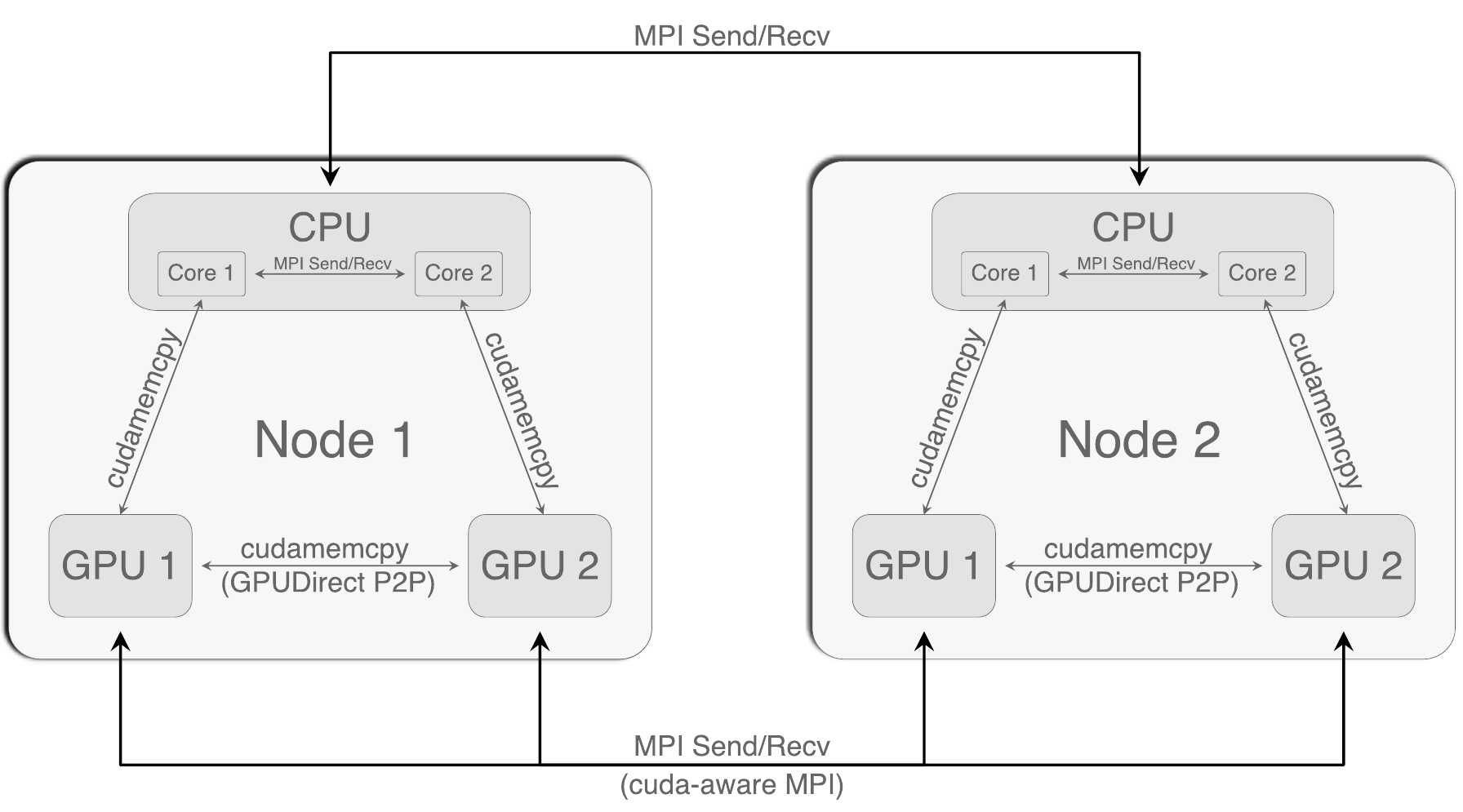}
\caption{{\it Left}: Example halo exchange process. Each process holds a sublattice,
         indicated by the gray square. Copies of sites located near the 
         borders of the sublattices, set off here by lines, are stored in the halos, 
         shown in white. {\it Right}:
         Diagram summarizing different possible communication channels for
         hardware.  Images taken from Ref.~\cite{mazur_topological_2021}.}
\label{fig:communicate}
\end{figure}

To work with multiple devices, $\simulat$ splits a lattice into multiple
sublattices, with partitioning possible along any of the four Euclidean
space-time directions. Each sublattice is given to a single GPU. In addition to
holding a field restricted to that sublattice, which we call the {\it bulk}, 
the GPU holds a copy of that field from the borders of the neighboring 
sublattices -- we call these copies the {\it halo}. A schematic drawing of the exchange of halos between different GPUs is shown in Fig.~\ref{fig:communicate} (left).

Communication between multiple CPUs and multiple nodes is handled with MPI, which also allows for communication between multiple GPUs. 
For NVIDIA hardware, communication via CUDA
GPUDirect P2P for intra-node and CUDA-aware MPI for inter-node channels is supported.
We boost performance by allowing the code to
carry out certain computations while communicating, such as copying halo buffers
into the bulk, whenever possible.

\begin{minipage}{\linewidth}
\begin{lstlisting}[captionpos=b,caption={Example plaquette kernel implemented as
a functor.},label={lst:plaquetteKernel},language=C++]
template<class floatT, bool onDevice, size_t HaloDepth>
struct CalcPlaq {
  gaugeAccessor<floatT> gaugeAccessor;
  CalcPlaq(Gaugefield<floatT,onDevice,HaloDepth> &gauge) : gaugeAccessor(gauge.getAccessor()){}
  __device__ __host__ floatT operator()(gSite site) {
    floatT result = 0;
    for (int nu = 1; nu < 4; nu++) {
      for (int mu = 0; mu < nu; mu++) {
        result += tr_d(gaugeAccessor.template getLinkPath<All, HaloDepth>(site, mu, nu, Back(mu), Back(nu)));
      }
    }
    return result;
  }
};

LatticeContainer<true, floatT> lattContainer(...);
Gaugefield<floatT, true, HaloDepth> gauge(...);
lattContainer.template iterateOverBulk<All, HaloDepth>(CalcPlaq<floatT, HaloDepth>(gauge))
\end{lstlisting}
\end{minipage}

This entire process by which splitting the lattice and communicating the halos
is carried out is quite technical and would generally make programming
operations running over all sites difficult to read, prone to mistakes,
and daunting to a newcomer. Therefore we have opted to abstract away indexing
and communication using {\it functor syntax}. Here, an operation that should be
performed at, for example, each lattice site is wrapped in the function call \texttt{operator()} 
of a \texttt{struct} or \texttt{class}. The functor takes as argument a
\texttt{gSite} object, which contains all relevant indexing information about a site
such as its coordinates. The functor is then passed to one of a few abstract launch functions 
that control on which sites the functor should be evaluated. These then create the necessary 
indexing objects and pass them, together with the functor, to the \texttt{RunFunctors} class, 
which finally performs the kernel launch.  
An example functor usage is shown in Listing~\ref{lst:plaquetteKernel}.
We have tried to make all code in $\simulat$ very general and therefore made
heavy use of templating. 
For instance, in Listing~\ref{lst:plaquetteKernel}, \texttt{floatT} is the 
precision, \texttt{onDevice=true} means we use
GPUs, \texttt{HaloDepth} determines the size of the halo,
and \texttt{Layout=All} means we do not split sites by even/odd parity.
We have suppressed the arguments of \texttt{lattContainer} and 
\texttt{gauge} for simplicity.

\begin{figure}
\centering
\includegraphics[width=0.85\textwidth]{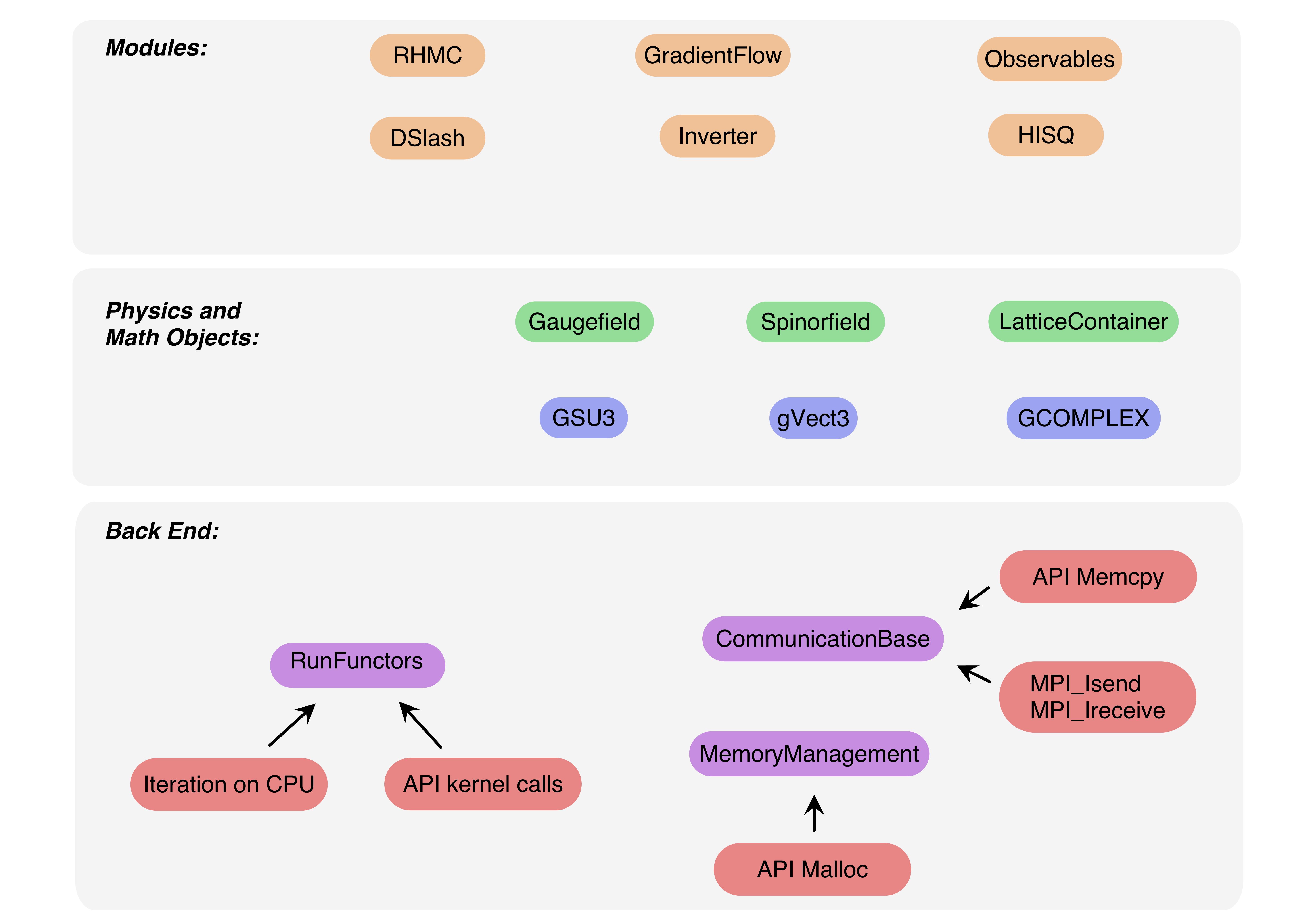}
\caption{Diagram illustrating the code's organizational hierarchy.}
\label{fig:organize}
\end{figure}

All of these lattice splitting, indexing, and communication procedures make up the foundation of our code, together with classes that manage the allocation of dynamic memory, file input/output, logging, and common math operations. Physics and
mathematics objects inherit from this backend and heavily utilize the functor
syntax. At the highest organizational level are the modules, which are
constructed from these physics and mathematics objects. At this highest level,
we strive to write code that closely and obviously mimics mathematical formulas
or short, descriptive English sentences. An overview of our code's
organizational scheme is depicted in Fig.~\ref{fig:organize}.
Adopting functor syntax and encapsulating our code according to this scheme has
already paid dividends. For example, while $\simulat$ was originally written for 
CUDA, porting our code to use HIP was accomplished by changing only few lines in the backend code,
and we find almost no performance difference
between the HIP and CUDA backends for NVIDIA GPUs.

Modules include, but are not limited to,
Wilson and Zeuthen flows~\cite{luscher_properties_2010,ramos_symanzik_2016};
heat bath~\cite{cabibbo_new_1982,kennedy_improved_1985} and 
over-relaxation~\cite{adler_over-relaxation_1981,creutz_overrelaxation_1987} 
updates for generating pure $\SU(3)$ configurations;
an RHMC~\cite{clark_rhmc_2004} for $N_f=2+1$ HISQ fermions;
gauge fixing via over-relaxation~\cite{mandula_efficient_1990};
sublattice updates~\cite{luscher_locality_2001,meyer_locality_2003}; and
3D and 4D all-to-all correlations for arbitrary functions of 
arbitrary operators. To carry out matrix inversion we use the conjugate gradient
(CG) methods, for which we have multiple right-hand side (multi-RHS), multiple shift, and mixed
precision implementations to improve performance.

\section{Performance}\label{sec:perform}

\begin{figure}
\centering
\includegraphics[width=0.7\textwidth]{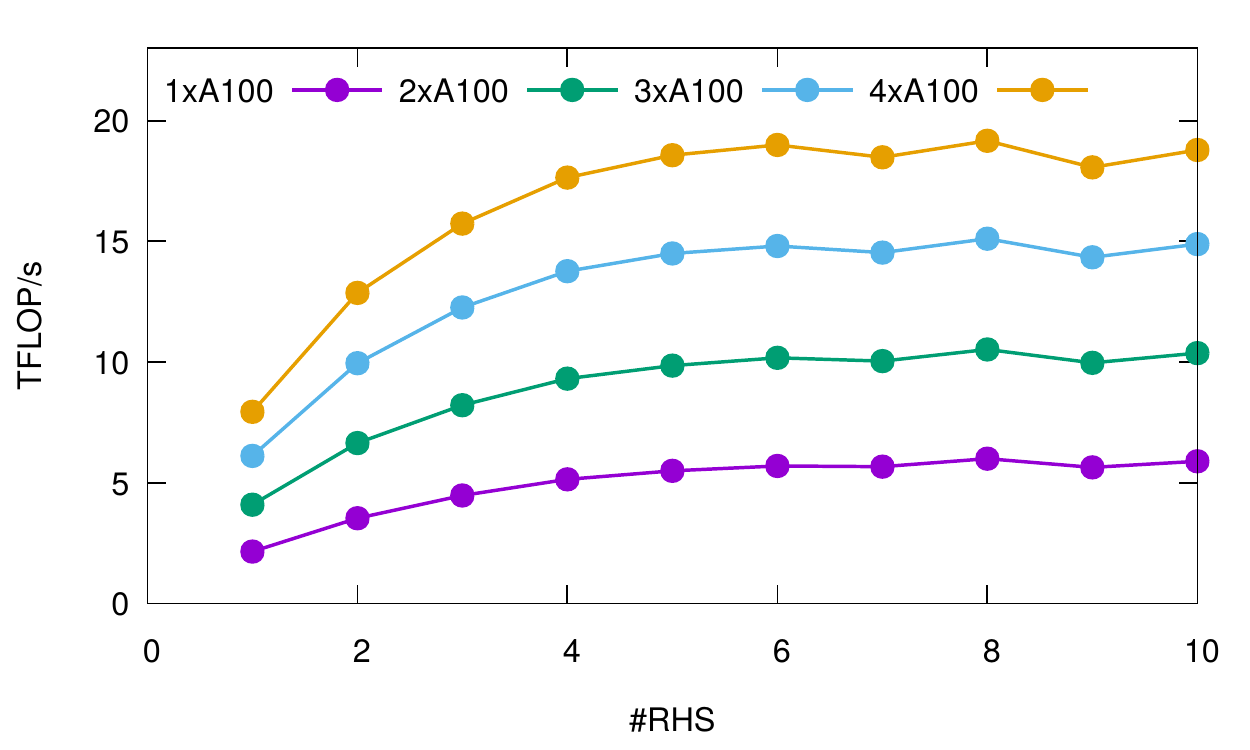}
\hspace{2mm}
\includegraphics[width=0.48\textwidth]{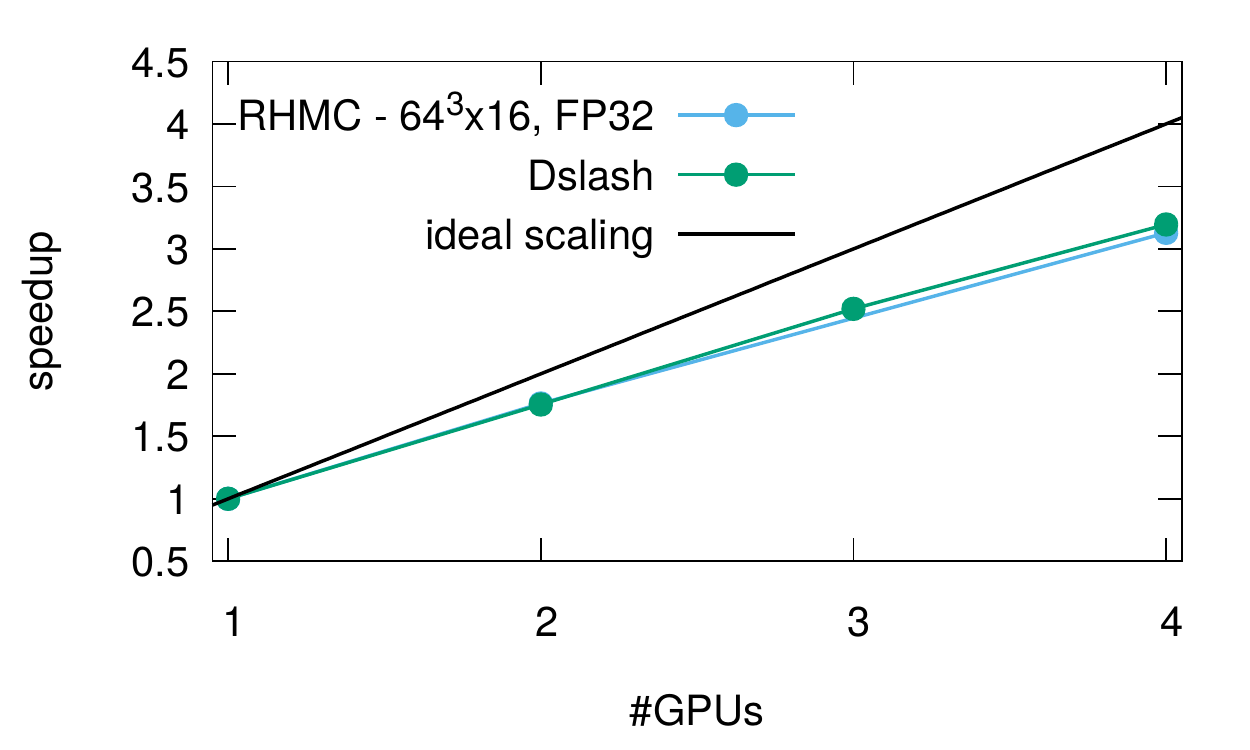}
\includegraphics[width=0.48\textwidth]{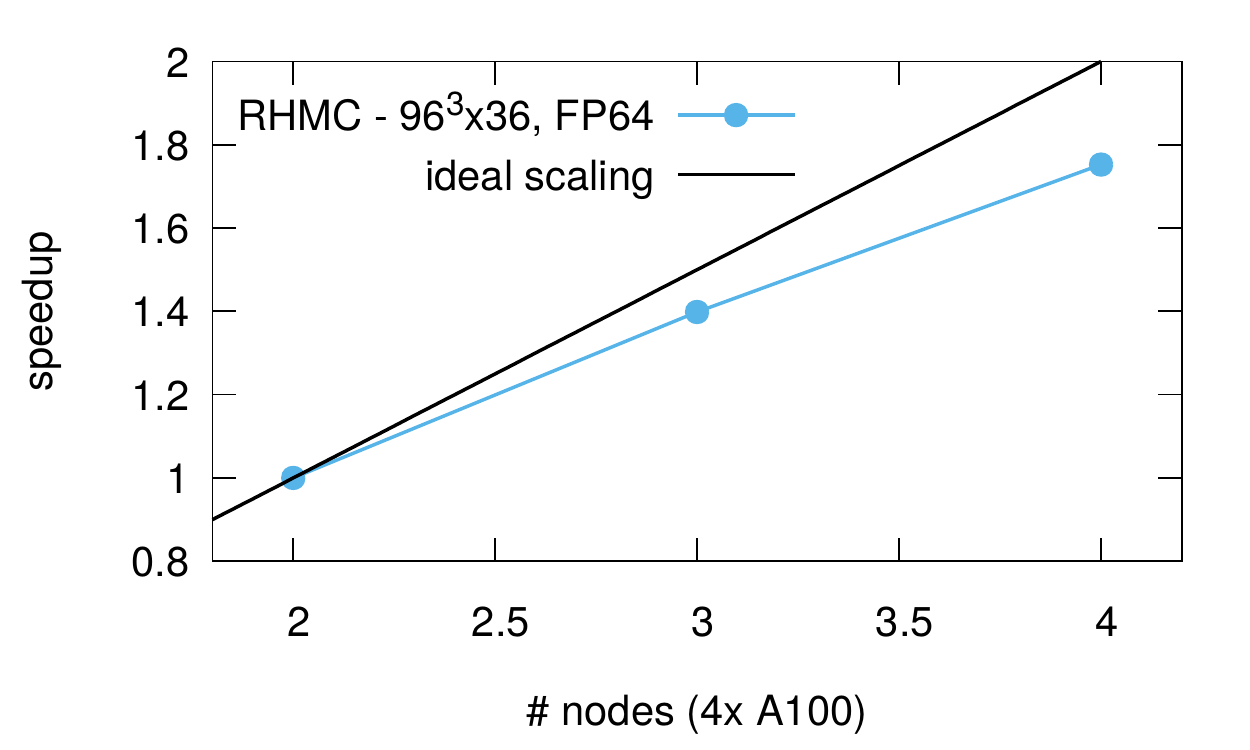}
\caption{{\it Top}: Scaling of multi-RHS $\slashed{D}$ with number of RHS
         for various numbers of GPUs on a single JUWELS Booster node. {\it Left}: Scaling of $\slashed{D}$, 
         and RHMC with number of GPUS on a single JUWELS Booster node. {\it Right}: Scaling of RHMC with multiple nodes with 4 A100 GPUs each.}
\label{fig:scaling}
\end{figure}

When generating HISQ configurations for typical parameters, about 60\% of our RHMC run time is spent
inverting the Dirac matrix via CG, and in it, applying the $\slashed{D}$ operator 
to a vector is the most performance-critical kernel.
Hence this and related kernels are good candidates for benchmarking. 
The kernel's performance is limited by the available memory bandwidth, but its arithmetic intensity can be increased by applying the 
gauge field to multiple RHS simultaneously. Furthermore, it benefits from gauge field compression. Only a subset of the link matrix entries are stored in memory, and the missing entries are recomputed from the stored ones based on the symmetries, either $\SU(3)$ or $\U(3)$, of the link matrix.
Figure~\ref{fig:scaling} (top) shows how the performance of the multi-RHS $\slashed{D}$
computed in single precision on a single JUWELS Booster node scales with number 
of RHS and GPUs. We achieve up to 1.36TB/s memory throughput and up to 19
TFLOP/s on one node. 
We also examine the speedup of the full RHMC algorithm on a single Booster node in Fig.~\ref{fig:scaling} (left) and compare it to the scaling of $\slashed{D}$. The scaling across a small number of nodes is also shown in Fig.~\ref{fig:scaling} (right).

\section{Conclusion and outlook}\label{sec:outlook}

We have presented $\simulat$, our multi-GPU, multi-node, high-performance
lattice code written in C++. 
Dynamical quarks are currently supported through the HISQ discretization, 
although other fermion discretizations can  be implemented.
We have adopted functor syntax, made heavy use of
templating, and organized our classes so that the code is easily adaptable to
new APIs, and so that lattice practitioners can easily write physics code  without needing a
thorough understanding of GPU parallelization. 
On multiple modern clusters, we find that its performance scales quite well with increasing
number of GPUs on a single node.
An open source release is forthcoming; it will be available in the near future in the repository 
linked in Ref.~\cite{github}.

\section*{Acknowledgements}

This work was supported by the Deutsche Forschungsgemeinschaft (DFG, German Research Foundation) 
Proj. No. 315477589-TRR 211.
This work was partly performed in the framework of the PUNCH4NFDI consortium supported by DFG fund "NFDI 39/1", Germany.
The benchmarks in this work were performed on JUWELS Booster and the GPU cluster at Bielefeld University.
We thank the Bielefeld HPC.NRW team for their support.

\bibliographystyle{JHEP}
\bibliography{bibliography}

\end{document}